# Ptychographic sensor for large-scale lensless microbial monitoring with high spatiotemporal resolution

Shaowei Jiang[1, 4], Chengfei Guo[1, 4*], Zichao Bian[1, 4], Ruihai Wang[1], Jiakai Zhu[1], Pengming Song[1], Patrick Hu[2], Derek Hu[3], Zibang Zhang[1], Kazunori Hoshino[1], Bin Feng[1], and Guoan Zheng[1,*]

[1]Department of Biomedical Engineering, University of Connecticut, Storrs, CT 06269, USA
[2]Department of Computer Science, University of California Irvine, Irvine, CA, 92697, USA
[3]Amador Valley High School, Pleasanton, CA, 94566, USA
[4]These authors contributed equally to this work.

*Correspondence:
C. G. (chengfei.guo@uconn.edu) or G. Z. (guoan.zheng@uconn.edu)

**Abstract**
Traditional microbial detection methods often rely on the overall property of microbial cultures and cannot resolve individual growth event at high spatiotemporal resolution. As a result, they require bacteria to grow to confluence and then interpret the results. Here, we demonstrate the application of an integrated ptychographic sensor for lensless cytometric analysis of microbial cultures over a large scale and with high spatiotemporal resolution. The reported device can be placed within a regular incubator or used as a standalone incubating unit for long-term microbial monitoring. For longitudinal study where massive data are acquired at sequential time points, we report a new temporal-similarity constraint to increase the temporal resolution of ptychographic reconstruction by 7-fold. With this strategy, the reported device achieves a centimeter-scale field of view, a half-pitch spatial resolution of 488 nm, and a temporal resolution of 15-second intervals. For the first time, we report the direct observation of bacterial growth in a 15-second interval by tracking the phase wraps of the recovered images, with high phase sensitivity like that in interferometric measurements. We also characterize cell growth via longitudinal dry mass measurement and perform rapid bacterial detection at low concentrations. For drug-screening application, we demonstrate proof-of-concept antibiotic susceptibility testing and perform single-cell analysis of antibiotic-induced filamentation. The combination of high phase sensitivity, high spatiotemporal resolution, and large field of view is unique among existing microscopy techniques. As a quantitative and miniaturized platform, it can improve studies with microorganisms and other biospecimens at resource-limited settings.

**Keywords:** temporal-similarity constraint, lensless microscopy, ptychography, quantitative phase imaging, dry mass measurement, filamentation.

**Highlights:**
▪ High sensitivity, centimeter-scale field of view, sub-micron spatial resolution
▪ Lensless monitoring of microbial growth at centimeter scale in 15-second interval
▪ Sensitive detection by tracking the phase wraps of the recovered images
▪ Quantitative dry mass measurement of bacterial cells in high throughput
▪ Platform technology for easy integration with other lab-on-a-chip devices

# 1. Introduction

Optical detection is an important tool for interrogating the growth and behavior mechanism of microorganisms. Performing rapid microbial detection via optical methods can shorten the time for antibiotic susceptibility testing and microbial limit testing, which are of critical importance in clinical, environmental, food, and pharmaceutical applications (Behera et al. 2019; Clontz 2008; Szczotka-Flynn et al. 2010). In the past years, different optical methods have been demonstrated for microbial detection and monitoring. For example, the VITEK system measures the light attenuation through the cultured wells and determines the growth conditions from the reading (Pincus 2006). The BacterioScan FLLS system uses forward laser light scattering to measure bacterial growth (Hayden et al. 2016). The Sensititre system detects the fluorescence signal for monitoring the activity of enzymes produced by microorganisms under testing (Luna et al. 2007). These and other traditional culture-based microbial detection methods, however, rely on the overall optical property of microbial cultures and cannot resolve individual growth event at single-cell resolution. As a result, they may require bacteria to grow to confluence and then interpret the results. Additionally, many laboratories are discovering that microorganisms, when stressed due to nutrient deprivation or following exposure to sub-lethal concentrations of antimicrobial agents, may not replicate when cultured on agar-based growth media (Justice et al. 2008). Antibiotic-induced filamentation is a good example where cells continue to elongate but do not divide. It protects bacteria from antibiotics and is associated with bacterial virulence such as biofilm formation. In a conventional culture-based experiment, it is challenging to detect elongated filamentous cells on the agar plate over a large field of view.

In contrast with the conventional microbial detection methods, image-based monitoring allows direct visualization of individual cells. Optical microscope has long been a standard tool for bacterial monitoring and cytometric analysis. However, the conventional microscope platform has the trade-off between resolution and imaging field of view. One can have a large field of view and a poor resolution, or a small field of view and a good resolution -- but not both. Therefore, it is challenging to achieve single-cell resolution of bacterial culture over a large field of view. Strategies to get both large field of view and high resolution have attracted much attention in recent years (Park et al. 2021). Fourier ptychographic microscopy (FPM) is one example that performs large field of view imaging with high resolution. It employs an LED array to illuminate the sample from different incident angles and perform aperture synthesizing in the Fourier domain for resolution improvement (Zheng et al. 2013). However, it may be challenging for FPM to image bacterial cells on agar plate. The main reason is that the agar substrate is not a transparent layer like glass slide and bacterial colonies are not confined to the 2D surface. The 2D aperture synthesizing process in FPM may not be valid for such a 3D object (Dong et al. 2014; Zheng et al. 2021; Zuo et al. 2020). Another strategy to expand the field of view of a regular microscope is to scan the sample as that in whole slide imaging systems. Such systems, however, are often expensive and require high maintenance. Robust and accurate autofocusing is a challenging task because of the micron-level short depth of field and centimeter-scale large sample size (Bian et al. 2020). Due to the 3D nature of bacterial colonies, it is also difficult to quantify the dry mass of all bacteria cells from 2D intensity images.

Miniaturization of imaging tools without lens can benefit rapid microbial detection and monitoring. In recognition of this need, low-cost lensless on-chip imaging techniques have been demonstrated for imaging various types of microorganisms, from yeast culture, bacteria, to waterborne parasites. These lensless techniques include optofluidic microscopy (Cui et al. 2008; Zheng et al. 2010), on-chip contact imaging (Jung and Lee 2016; Lee et al. 2014; Lee et al. 2012; Zheng et al. 2011), digital in-line holography (Mudanyali et al. 2010; Su et al. 2010; Wang et al. 2020; Xu et al. 2001), multi-height and multi-wavelength imaging (Bao et al. 2008; Greenbaum and Ozcan 2012; Greenbaum et al. 2014; Latychevskaia 2019; Luo et al. 2016; Wu et al. 2021), near-field ptychographic structured modulation microscopy (Jiang et al. 2020), among others. For example, a contact-imaging-based digital Petri dish platform (ePetri) has been used to perform long-term imaging of motile microorganisms and bacterial cultures (Jung and Lee 2016; Lee et al. 2012). It was shown that time-series images of *S. epidermindis* with single-cell resolution can be obtained from an integrated system and the results are comparable to conventional colony-counting assays (Jung and Lee 2016). Similarly, digital in-line holography has also been demonstrated for imaging bacterial culture over a large field of view with a half-pitch resolution of 3.5 µm (Wang et al. 2020). Despite the success of these previous demonstrations, the captured images from these platforms cannot be used as a quantitative measure of microbial growth. For contact-imaging methods, the captured 2D intensity image can only be used to determine the 2D area occupied by the 3D bacterial colonies. Culturing on the surface of the image sensor also differs from the workflow of conventional methods. For digital in-line holography and multi-height phase retrieval, it is challenging to recover the true quantitative phase of 3D colonies on uneven agar plate, which contains slow-varying phase information with many $2\pi$ wraps. Similar to the contact imaging methods, they can only track the 2D area occupied by large 3D bacterial colonies from the intensity images. It is challenging for them to recover the correct $2\pi$ wraps for tracking bacterial growth with high phase sensitivity.

Here, we demonstrate the development of an integrated ptychographic sensor for lensless cytometric analysis of microbial cultures over a large scale and with a high spatiotemporal resolution. Our device integrates the concepts of the digital Petri dish platform (Jung and Lee 2016) and ptychographic structured modulation microscopy (Jiang et al. 2020; Song et al. 2019). Similar to the digital Petri dish platform, the integrated ptychographic sensor can be placed within a regular incubator or used as a standalone incubating unit for long-term microbial monitoring. To acquire the image of live bacterial cultures, we can translate the ptychographic sensor under the agar plate for lensless data acquisition. The acquired data from the ptychographic sensor is then used to recover the complex exit waves diffracted by the specimen, which can be further refocused to any plane along the axial direction post measurement. For longitudinal study where massive data are acquired at sequential time points, we report a new temporal-similarity constraint to increase the temporal resolution of ptychographic reconstruction by at least 7-fold. With this strategy, the reported device achieves a centimeter-scale field of view, a half-pitch spatial resolution of 488 nm, and a temporal resolution of 15-second intervals. For the first time, we report the direct observation of bacterial growth in a 15-second interval by tracking the phase wraps of the recovered images. Without involving any interferometric measurement, the phase sensitivity of the reported low-cost device is similar to those obtained using high-end interferometric setups while achieving orders of magnitude higher throughput. The combination of high phase sensitivity, high spatiotemporal resolution, and large field of view is also unique among existing microscopy techniques. As a quantitative and miniaturized imaging platform, the reported technology can greatly improve studies and experiments with microorganisms and other biospecimens over a large scale. The temporal-similarity constraint also enables large-scale longitudinal ptychographic imaging with an unprecedented temporal resolution.

## 2. Materials and methods

### 2.1. Integrated ptychographic sensor

As shown in Fig. 1a, the reported ptychographic sensor takes a very simple geometry: an image sensor, a thermoelectric cooler, a heatsink, and a fan are integrated and placed under the specimen. A dense thin layer of 1-2 µm microspheres is directly coated on top of the sensor's coverglass (black dash line in Fig. 1a).

This scattering layer serves as a computational lens for improving the numerical aperture of the imaging platform (Choi et al. 2013; Song et al. 2019).

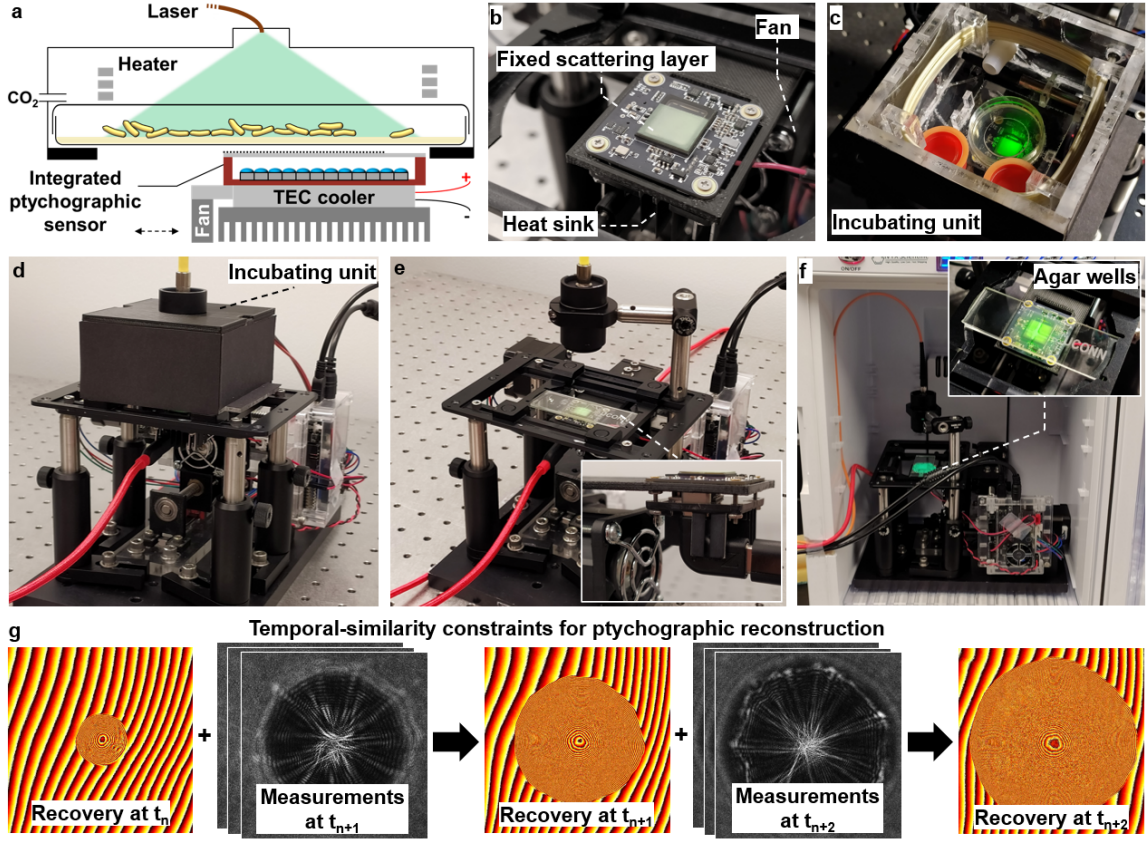

**Fig. 1. Integrated ptychographic sensor for large-scale lensless microbial monitoring with high spatiotemporal resolution.** (a) Operation principle of the integrated ptychographic sensor, which consists of an image sensor coated with a scattering layer, a TEC cooler, a heatsink, and a fan. By translating the integrated sensor under the specimen, we can acquire lensless data for ptychographic reconstruction. (b) The prototype of the integrated ptychographic sensor with the scattering layer coated on the sensor coverglass. (c) A mini-incubating unit for standalone application. (d) The self-incubating lensless imaging platform. (e)-(f) The imaging platform without the incubating unit can be placed within a regular incubator for image acquisition. (g) The temporal-similarity constraint for longitudinal study. The recovered image of bacterial colonies from the current time point is used as the initial guess of the next time point. With this strategy, we can increase the temporal resolution of ptychographic reconstruction by at least 7-fold.

To operate the system in Fig. 1a, we simply translate the integrated ptychographic sensor under the specimen and acquire the lensless images for reconstruction. Figure 1b shows our prototype device where the integrated sensor is built using a monochromatic image sensor with a pixel size of 1.85 µm (Sony IMX226). The integrated sensor is mounted on two translation stages modified from a low-cost computer numerical control router (Guo et al. 2020). The step size in-between adjacent acquisitions is 1-4 µm. Different from LED-based illumination demonstrated in lensless setups (Jung and Lee 2016; Lee et al. 2014; Lee et al. 2012; Mudanyali et al. 2010; Wang et al. 2020; Zheng et al. 2011), the use of 532-nm laser diode in our setup enables a short exposure time of <0.2 ms for image acquisition, allowing the sample to be in continuous motion without motion blur issues. In Fig. 1c, we also build a mini-incubating unit that can be coupled with the ptychographic sensor for standalone operation. In this mini-incubating unit, we use a heating resistor wire and a temperature sensor for feedback temperature control. Water is added to the two yellow containers to increase the humidity of the environment. The white tube in Fig. 1c can also be connected to a portable $CO_2$ supply. Figure 1d shows the entire compact imaging platform with the self-incubating unit. Similar to the digital Petri dish platform (Jung and Lee 2016; Lee et al. 2012), our compact and cost-effective device can also be placed within a regular incubator and stream the data via the USB 3.0 cable (Figs. 1e and 1f).

At the heart of our image recovery process is a coherent diffraction imaging (CDI) technique termed ptychography. This technique was original proposed to address the phase problem of crystallography in electron microscopy (Hoppe and Strube 1969) and brought to its modern form with an iterative phase retrieval framework (Faulkner and Rodenburg 2004). In a typical implementation, the specimen is laterally translated through a spatially confined probe beam. The lensless diffraction patterns, termed ptychograms, are then recorded at the reciprocal space using a 2D image sensor. The reconstruction process iteratively imposes two different sets of constraints. The diffraction measurements serve as the Fourier magnitude constraints in the reciprocal space. The confined ptychographic probe beam limits the physical extent of the object for each measurement and serves as the support constraint in the real space.

For the reported ptychographic sensor in Fig. 1a, we illuminate the entire sample with an unconfined laser beam and use the scattering layer on the sensor coverglass as an unconfined ptychographic probe for encoding the object exit waves into diffraction measurements (Song et al. 2019). The precise estimation of positional shifts $x_j$ and $y_j$ of the ptychographic sensor is important for achieving high resolution in reconstruction. The continuous motion of the sample and the open-loop operation of the motorized stage also prevent us to precisely retrieve the scanning positions of the ptychographic sensor. To address this problem, we remove the scattering layer from the part of the sensor's coverglass in Figs. 1a-1b. We then use the first image captured through this clear region as the reference image. For all subsequent measurements, we identify the translational shift by locating the maximum point of the cross-correlation between the reference image and the subsequent measurements (Bian et al. 2019).

With the recovered translation shift, we can then employ a calibration object (blood smear or histology slide) to characterize the diffusing layer (Song et al. 2019), using a procedure similar to the joint object-probe updating process of blind ptychography (Guizar-Sicairos and Fienup 2008; Maiden and Rodenburg 2009; Ou et al. 2014; Thibault et al. 2009). The complex profiles of bacterial colonies in subsequent experiments can then be recovered using a ptychographic phase retrieval process similar to those in (Bian et al. 2019; Jiang et al. 2020; Song et al. 2019). The detailed recovery process is provided in Supplementary Note 1. We note that the 1.85 µm pixel size of the employed image sensor is not a limiting factor for the final achievable resolution. In the recovery process, we up-sample the reconstruction by $M$-fold. As such, each raw pixel of the detector corresponds to $M$ by $M$ subpixels in the reconstruction. We then enforce the intensity summation of every $M$ by $M$ subpixels equals the corresponding raw pixel in the captured image (Batey et al. 2014; Guo et al. 2016; Jiang et al. 2020). With this strategy, we can resolve 488 nm linewidth on the resolution target using the sensor with 1.85-µm pixel size (discussed in the Results section).

**2.2. Temporal-similarity constraint for improving spatiotemporal resolution**

A robust phase retrieval often requires the use of more measurements or other prior object knowledge to properly constrain the solution space. In digital in-line holography, compact object support constraints can be imposed in the iterative reconstruction process for imaging sparse samples (Fienup 1987; Miao et al. 1999; Mudanyali et al. 2010; Su et al. 2010). In this case, a threshold or a segmentation algorithm can be used to estimate the objects' locations and create a mask for the object support. Within the support, the object will be updated using the diffraction measurements; outside the support, the signal will be set as background. For conventional real-space ptychography (Faulkner and Rodenburg 2004), the confined probe beam defines a compact support for the phase retrieval process. Similarly, the confined aperture of the objective lens defines a compact support in the reciprocal space for FPM (Zheng et al. 2013).

Here, we report a new type of constraint, termed temporal-similarity constraint, for significantly accelerating the phase retrieval process. The key observation of longitudinal study is that the recovered object images are similar when captured at adjacent time points. In the proposed constraint (Fig. 1g), we use the recovered image from the last time point as the initial guess for the current time point. This constraint, essentially, enforces a good initial guess for the phase retrieval process. As we will discuss in a later section, ~70 raw images can generate a high-quality reconstruction of bacterial colonies. In comparison, ~490 images are needed to achieve a similar reconstruction quality if this constraint is not imposed. By imposing this constraint, the recovery process can also be shortened thanks to the reduction of the captured images.

The strategy of imposing temporal-similarity constraints can also be extended to objects with slow continuous motion, such as the motile microorganisms demonstrated in (Lee et al. 2012). Different from the confined probe in conventional ptychography, we use the scattering layer as an unconfined probe for information encoding. Each measurement of our device, thus, contains information of the entire sample with a large field of view. We can perform ptychographic reconstruction at camera framerate using the temporal similarity constraint. For regular ptychography and FPM, the captured images only contain the encoded information from a localized probe. To update the entire object, we still need to perform a full scan in either the spatial or the Fourier domain.

**2.3. Dry mass measurement of cells**

In the reported device, the optical phase shift accumulated through a live cell is linearly proportional to the dry mass (nonaqueous content) of the cell. The dry mass density at each pixel can be calculated as $\rho(x,y) = (\lambda/2\pi\gamma)\phi(x,y)$, where $\lambda$ is the center wavelength, $\gamma$ is the average refractive increment of protein (0.2 mL/g), and $\phi(x,y)$ is the measured phase (Mir et al. 2011). The total dry mass can then be obtained by integrating the density over the region of interest. Dry mass measurement provides a quantitative metric for characterizing cell growth. The unique capability of dry mass tracking over a large area with single-cell resolution, therefore, allows us to perform rapid bacterial detection at low concentrations, which is critical for medical, environmental, and food safety applications.

**3. Results**

**3.1. Resolution performance and quantitative phase imaging**

We first validate the imaging performance using a resolution target and a quantitative phase target. Figures 2a and 2b show the raw measurement and the recovered image of the resolution target. In the recovered image, we can solve group 10, element 1 on the target and the corresponding linewidth is 488 nm. In contrast, the pixel size is 1.85 µm in the raw image. Supplementary Fig. S1 further shows the resolution performance using different scattering layers. To achieve high resolution in the reported device, the key consideration is to use a dense but thin scattering layer on the sensor coverglass. Supplementary Fig. S2 shows the resolution performance using different number of measurements. We also note that the 488-nm half-pitch resolution demonstrated here is achieved without aperture synthesizing. To further improve the resolution, it is possible to use angle-varied illumination for aperture synthesizing. However, the aperture synthesizing process may only be valid for 2D thin objects. The 3D nature of bacterial colonies may pose a challenge for the 2D aperture synthesizing process (Zheng et al. 2021; Zuo et al. 2020).

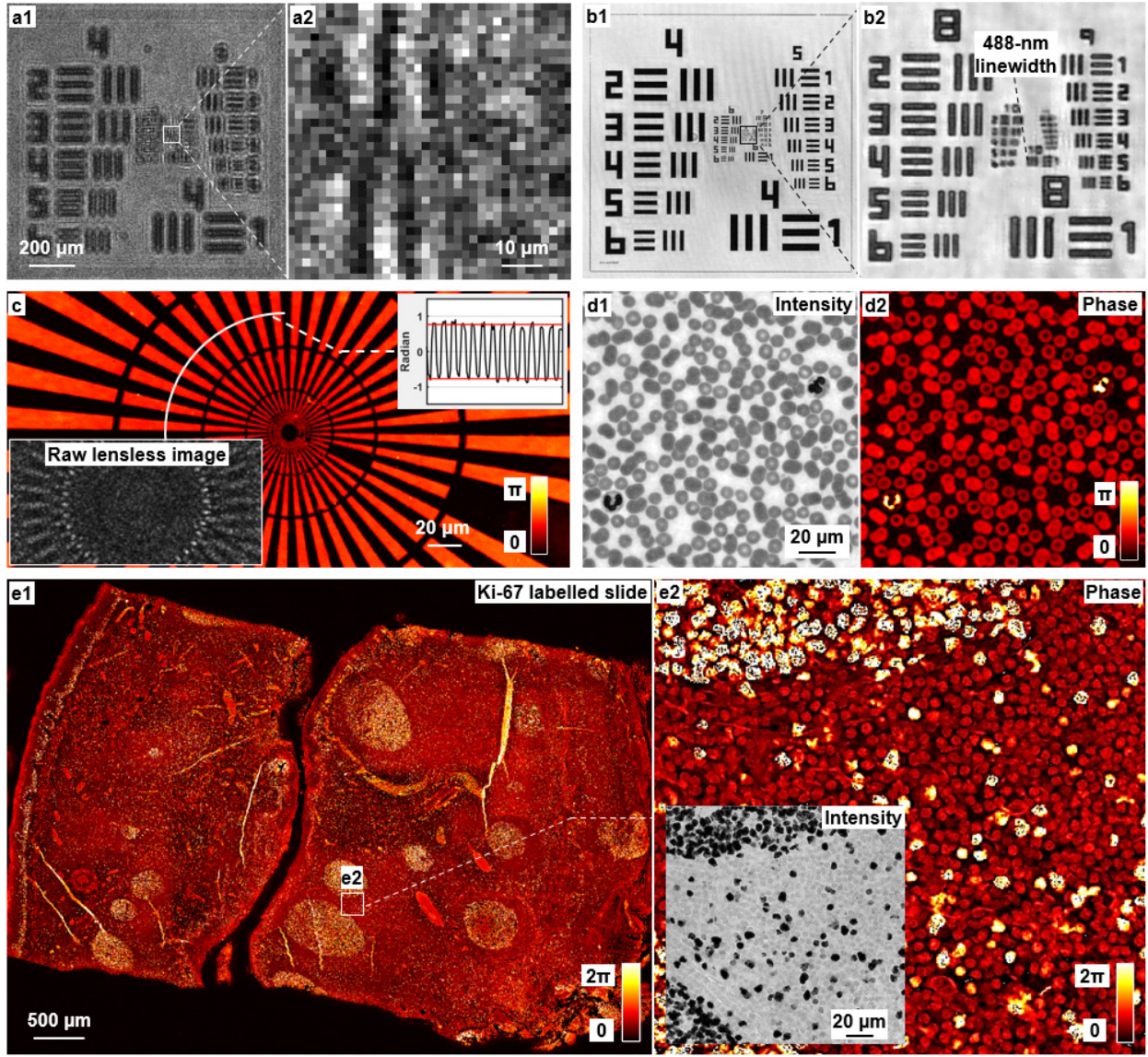

**Fig. 2. Imaging performance of the integrated ptychographic sensor.** (a) The captured raw image of a resolution target, with a raw pixel size of 1.85 μm. (b) The recovered image of the resolution target, where we can resolve 488-nm linewidth of group 10, element 1. (c) The recovered phase of the phase target. Left bottom inset shows the raw image. Top right inset shows the line trace of the phase. (d) The recovered intensity and phase images of a blood smear. (e) The recovered whole slide phase image of IHC slide labeled with Ki-67 biomarker. The whole slide phase image is shown in (e1) and the magnified view is shown in (e2). Inset of (e2) shows the corresponding recovered intensity of the IHC slide.

In Fig. 2c, we show the recovered phase of the quantitative phase target. The inset on the left bottom shows the corresponding raw image and the inset on the top right shows the line trace of the phase. We can see that the recovered phase is in good agreement with the ground truth (red line in the inset). In Fig. 2d1 and 2d2, we show the recovered intensity and phase images of a blood smear (Carolina Biological Supply). To demonstrate the large field of view capability of the reported device, we show the recovered whole slide phase image of an immunohistochemistry (IHC) section in Fig. 2e, where the cells are labeled by the Ki-67 biomarker ([Matsuta et al. 1996](#)). In Supplementary Fig. S3, we validate the imaging performance of our device by comparing our results with those captured using a regular light microscope. In the demonstrations shown in Fig. 2, we acquire 450 images in 15 seconds for reconstruction. Supplementary Fig. S4 further shows the recovered blood smear with different number of measurements.

**3.2. Imaging with high sensitivity, centimeter-scale field of view, and high spatiotemporal resolution**

For imaging live bacteria, one challenge is the slow varying phase profile with a large phase range from the bacterial culture. In common lensless techniques such as digital in-line holography and contact imaging, the slow-varying phase profiles of bacterial colonies cannot be effectively converted into intensity variations for detection. As such, the recovered phase may not be quantitative for these objects. Phase objects with a large phase range (with many 2π wraps) are also a general challenge to the phase retrieval process. It often requires more diversity measurements to better constrain the solution space.

Our solution for imaging live bacteria colonies with large phase range is to impose the temporal-similarity constraint outlined in Fig. 1g. In our experiment, 10 mL fresh Mueller-Hinton broth was first inoculated with a single colony of *E. coli* ATCC 25922 strain from the Mueller-Hinton agar plate and kept at 37 °C. The culture was then incubated in a culture tube at 37 °C overnight. On the following day, we adjusted the turbidity of the bacterial solution to 0.5 McFarland standard with fresh Mueller-Hinton medium, containing ~$10^8$ CFU / mL. The bacteria suspension was then diluted to a concentration of ~$10^3$ CFU / mL. The concentration of the final diluted bacteria suspension was also checked using the standard plate count method. To image live bacterial cultures, we added the prepared bacteria suspension to a regular Petri dish with Mueller-Hinton agar (Fig. 1c). The entire Petri dish was placed within our mini-incubating unit shown in Fig. 1d.

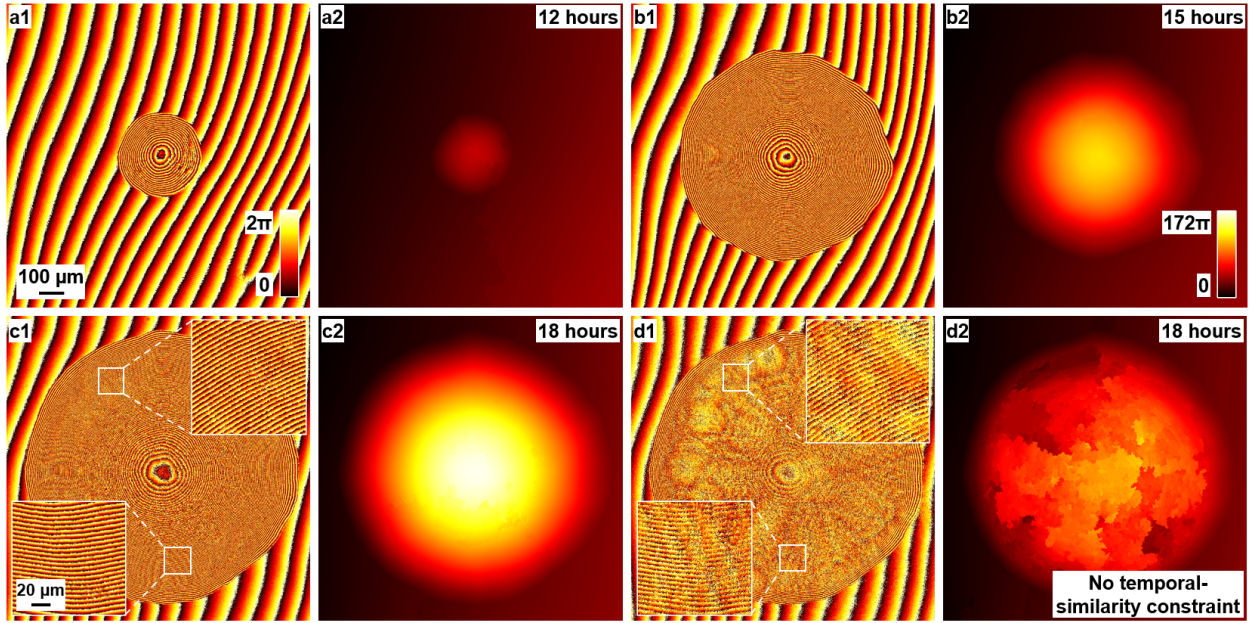

**Fig. 3. Temporal similarity constraint for imaging bacterial colonies.** (a1)-(c1) The recovered phase images of a live bacterial colony at different time points. 70 raw images were used for reconstruction at each time point. The temporal similarity constraint was imposed and it can significantly increase the temporal resolution by at least 7-fold. (a2)-(c2) The unwrapped phase of (a1)-(c1). (d1) The recovered phase image with 250 raw measurements. No temporal-similarity constraint is imposed as a comparison. (d2) The unwrapped phase of (d1). This demonstration validates the effectiveness of the reported temporal-similarity constraint.

Figure 3a-3c show the recovered and unwrapped phase images of the bacterial colony at 3 different time points. The reported temporal-similarity constraint is imposed for the ptychographic reconstruction. With this constraint, we can significantly reduce the number of acquisitions. In Fig. 3a-3c, we only need to use ~70 images for reconstruction. In contrast, Fig. 3d shows the reconstruction without using the temporal-similarity constraint. In this comparison, we use 250 raw images for reconstruction and the image quality is worse than those shown in Fig. 3a-3c. Supplementary Fig. S5 further compares the reconstructions with and without imposing the temporal-similarity constraint. We can see that the reported temporal-similarity constraint can significantly shorten the acquisition time and increase the temporal resolution of ptychographic reconstruction by at least 7-fold.

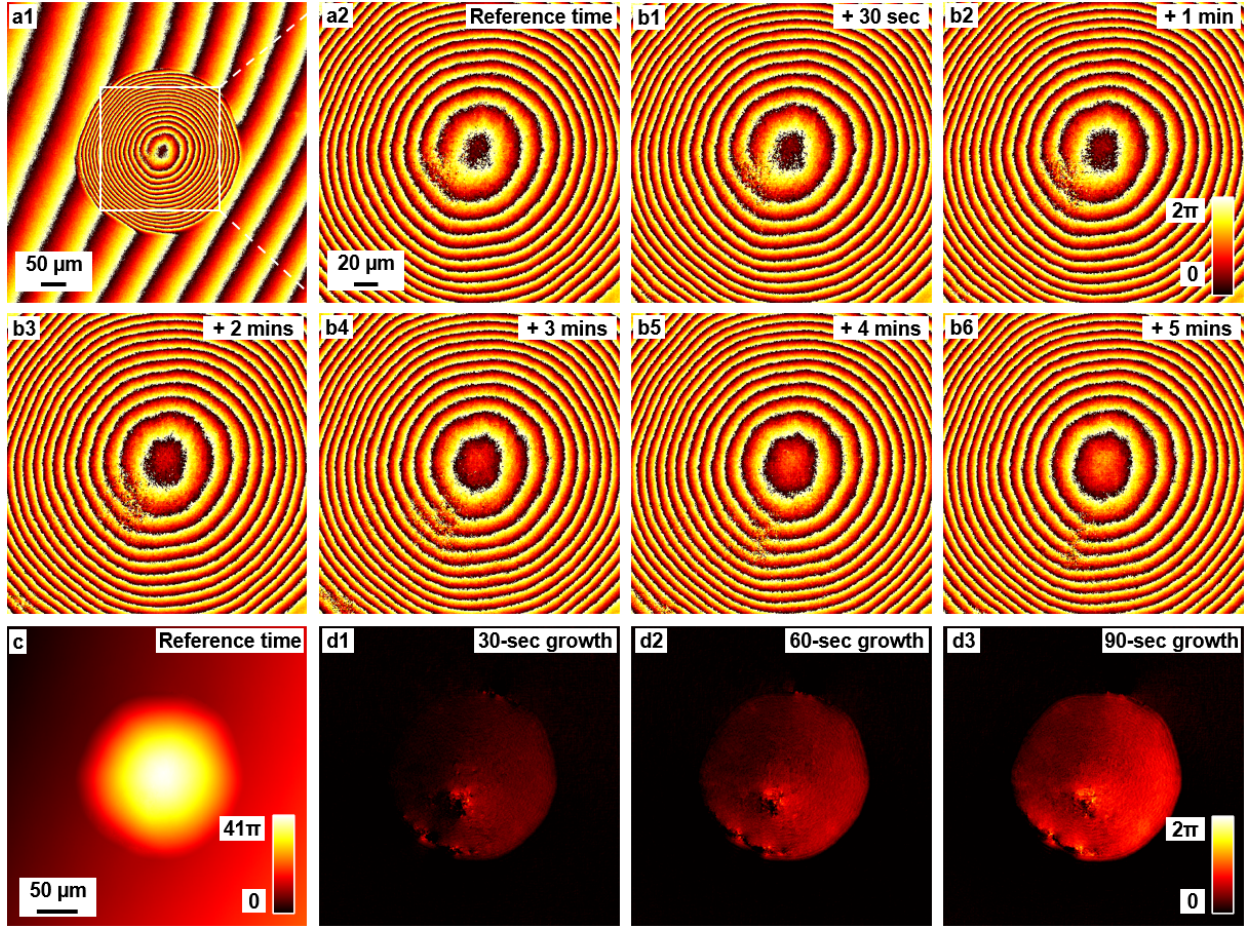

**Fig. 4. Sensitive detection of bacterial growth at high spatiotemporal resolution and with a centimeter-scale field of view.** (a1) The recovered phase image at the reference time point. (a2) The magnified view of (a1). (b1)-(b6) The recovered phase images at subsequent time points. We can image a centimeter-scale sample at a 15-second interval (also refer to Supplementary Video S1). (c) The unwrapped phase image of (a1). (d) The growth of the bacterial culture is visualized by the time-differential phase maps (the difference between the unwrapped phase maps at the reference and the subsequent time points). (d1)-(d3) shows the bacterial growth at 30-s, 60-s, and 90-s intervals.

Sensitive detection of bacterial growth is important for clinical, environmental, and pharmaceutical applications. Most current imaging solutions can only track the change of area occupied by the bacterial cells. The reported ptychographic sensor, on the other hand, can perform sensitive growth detection by tracking the phase wraps of the recovered image. Without involving any interferometric measurement, the phase sensitivity of our low-cost device is similar to those obtained using high-end interferometric setups. By using the temporal-similarity constraint, we can image 4 sensor fields of view in 15 seconds and the corresponding imaging area is ~120 mm$^2$ (Supplementary Fig. S6). For each field of view of the detector (~30 mm$^2$), we acquire ~70 images in 3 seconds. The stage then switches to the next field of view to repeat the acquisition process. Supplementary Video S1 shows the recovered phase images of the bacterial colony acquired at a 15-second interval. In this video, we also compare the reconstructions with (left) and without (right) the temporal-similarity constraint. Figure 4a shows the recovered phase at the reference time point and Fig. 4b shows the subsequent recovered images. For the first time, we can directly observe bacterial growth in a 15-second interval by tracking the phase wraps of the recovered images, achieving sensitive detection similar to that of interferometric measurements. We further virtualize the bacterial growth in Fig. 4c-4d. Figure 4c shows the unwrapped phase at the reference time point and Fig. 4d shows the time-differential phase maps of the subsequent measurements. The growth of colonies can be clearly visualized at 30-s, 60-s, and 90-s intervals shown in Fig. 4d.

### 3.3. Tracking of 3D bacterial colonies and quantitative dry mass measurement

For current microscopy techniques, observing bacterial growth transition from 2D single layer to 3D colony formation remains a challenge due to the low spatial frequency content of the layered structure. Confocal microscopy can be used to image the layered colony structure at a small region and with a limited axial resolution (Su et al. 2012). In Fig. 5a-5b, we can directly observe the forming of layered structures in microcolonies using the recovered phase images. The projected line traces in Fig. 5b show two-layer and three-layer structures, with ~0.5 radians of phase accumulation for each layer. The multilayer forming process occurs at the center while the monolayer retains in the outer regions. In this experiment, we subtract the background phase captured at time zero to better visualize the formation process of 3D micro-colony.

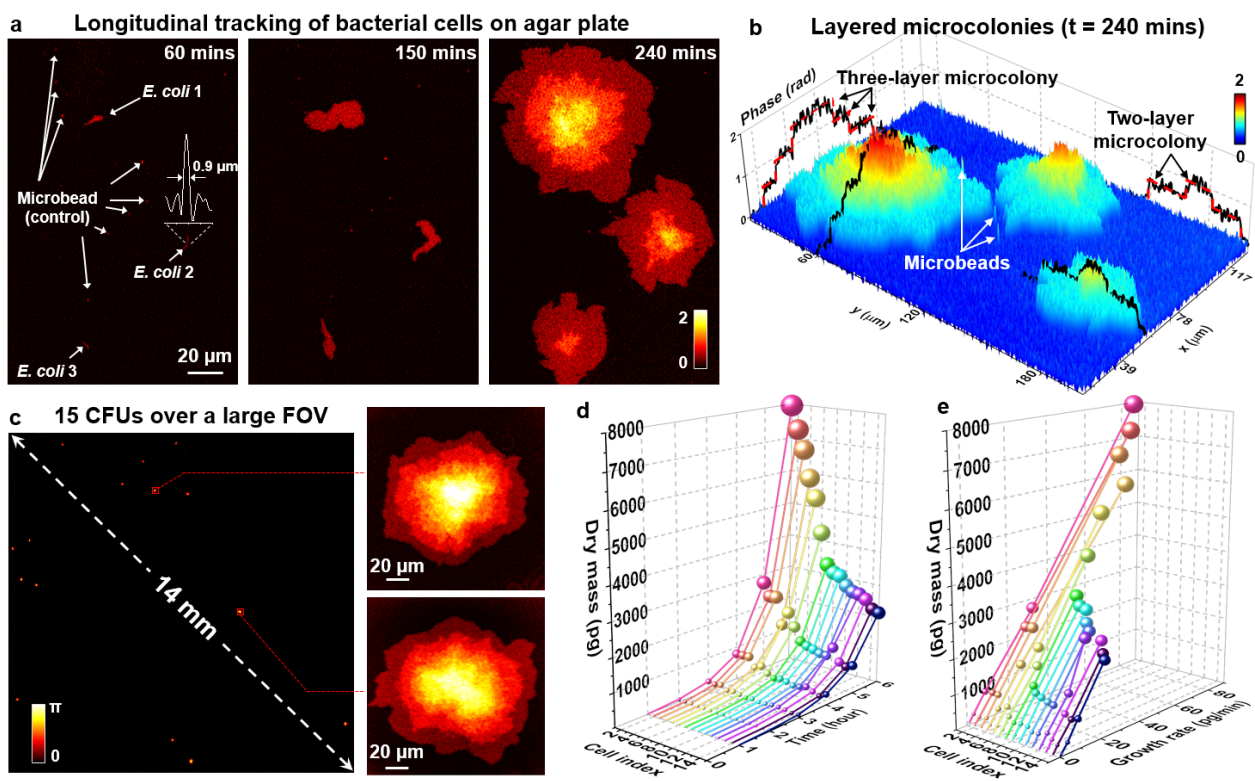

**Fig. 5. Tracking of 3D bacterial colonies and quantitative dry mass measurement.** (a) Recovered phase images of bacterial cells at different time points. Microbeads are used as a negative control. (b) The recovered phase map reveals the layered structure of 3D bacterial colonies. (c) Rapid bioburden testing for measuring the total number of viable microorganisms over the agar plate. 15 microcolonies are tracked over a centimeter-scale area. (d) The dry masses of the 15 microcolonies are tracked at different time points. (e) The dry masses of the 15 microcolonies are plotted as a function of growth rate.

Understanding how a bacterial cell becomes a 3D colony is important in biomedicine and food safety. Much is known about the molecular and genetic bases of the cells but less about the dynamic physical process. It has been shown that bacterial cells self-organize into high-density colonies in a pattern similar to the collective motions commonly seen with fish schools and flocking birds (Zhang et al. 2010). The reported device provides a turnkey tool for studying the growth of bacterial colonies in 3D, for which there is very little data at present. A predictive understanding of bacterial colony formation will constitute an advance in active matter physics and provides new insights for preventing the formation of tough biofilms (Qi et al. 2016).

In Fig. 5c, we use the reported platform for rapid bioburden testing that measures the total number of viable microorganisms of a sample. In this experiment, we apply 8 μL of *E. coli* suspension onto the agar plate and the suspension has a low concentration of ~1000 colony-forming units (CFUs) / mL. Figure 5c shows the recovered phase image where we track 15 CFUs on a ~14 mm agar plate. The time-lapse images of these 15 CFUs are shown in Supplementary Fig. S7. Figure 5d shows the longitudinal tracking of the dry mass for 15 microcolonies. Figure 5e shows the dry mass plot as a function of grow rates, where the linear relation indicates the exponential growth. The difference between dry mass tracking and cell area tracking is further provided in Supplementary Fig. S8, where the latter shows a ~60% error in quantifying the cell growth. Different from the conventional approaches, our platform can detect cell growth as soon as the dry mass increases. We can, thus, confirm the bacterial growth with low concentrations in hours instead of days. The detection upper limit of our current system is on the order of 100 CFUs / ml, assuming we identify one CFU over the 14 mm agar plate. As a reference, clinical infection is typically diagnosed with a concentration of ~$10^4$ CFU / mL. We can further improve the detection limit by imaging a larger field of view.

**3.4. Drug screening**

Rapid dry mass metrology can also be used for antibiotic susceptibility testing (AST) (Reller et al. 2009), which determines whether an antibiotic drug will be effective in stopping the growth of a specific bacterial strain. If an effective antibiotic can be administered in the early stages of infection, it can avert the development of antibiotic resistance in a clinical setting. In the experiment shown in Fig. 6, we add different amounts of gentamicin to 4 agar wells and perform average dry mass measurements of the bacterial colonies. Figure 6a shows the recovered phase images under different conditions. Figure 6b shows the average dry mass measurements of the cells over time. The minimum inhibitory concentration (MIC) value is determined to be ~1 μg / ml for the employed *E. coli* strain in 2 hours. The MIC value obtained using our device is at the same range as that determined by the reference microdilution method of the Clinical and Laboratory Standards Institute (CLSI 2021). In contrast, the standard culture-based approach often takes 24 hours to obtain the MIC result, a time during which a patient's clinical course can rapidly deteriorate.

A more interesting case for AST is shown in Fig. 6c, where we apply 8 μg / ml of ampicillin to the agar well for imaging. In this experiment, we observe the antibiotic-induced filamentation where cells continue to elongate but do not divide. Filamentation protects bacteria from antibiotics and is associated with bacterial virulence such as biofilm formation. Figure 6d tracks the dry mass and length of individual cells, where each spherical point represents the measurement of one filamentous cell with color coded for time. In a conventional culture-based AST, it is challenging to locate filamentous cells on the agar plate over a large field of view. Therefore, the applied drug may be incorrectly concluded as effective in stopping bacterial growth. The tracking of filamentous cells using the reported platform, on the other hand, enables quantitative measurement of bacterial growth in single-cell resolution and provides accurate information of drug susceptibility that cannot be obtained using the conventional approach.

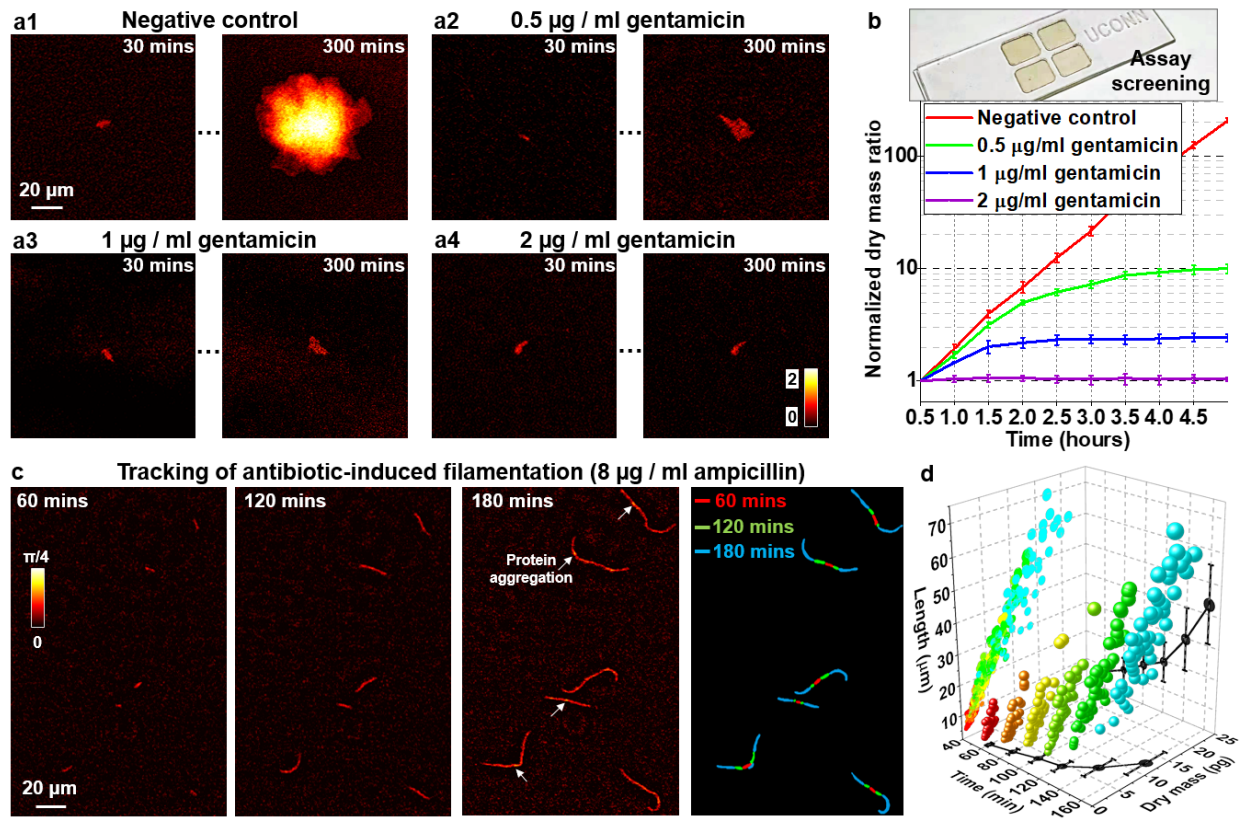

**Fig. 6. Rapid antibiotic drug screening**. (a) Different amounts of gentamicin are added to the 4 agar wells and the recovered phase images of bacterial cells are shown at different time points and with different drug treatments. (b) The average dry mass measurements at different wells are plotted as a function of time. The minimum inhibitory concentration value of ~1 μg / ml can be determined in 2 hours. (c)-(d) Tracking of antibiotic-induced filamentation of bacterial cells, where (c) shows the recovered phase with arrow highlighting protein aggregation. Each spherical point in (d) represents one filamentous cell and we measure its length and dry mass at different time points.

## 4. Discussion and conclusion

In summary, we have developed an integrated ptychographic sensor for lensless cytometric analysis of microbial cultures over a large scale and with high spatiotemporal resolution. We also report a new temporal-similarity constraint to increase the temporal resolution of ptychographic reconstruction by 7-fold. With this strategy, the reported device achieves a centimeter-scale field of view, a half-pitch spatial resolution of 488 nm, and a temporal resolution of 15-second intervals. Our platform is also highly sensitive to the phase change caused by bacterial growth. Without involving any interferometric measurement, the phase sensitivity of our device is similar to those obtained using high-end interferometric setups. With the high sensitivity to the phase, we report the direct observation of bacterial growth in a 15-second interval. Our demonstrations on dry mass measurement and drug screening indicate that the integrated ptychographic sensor can be a turnkey tool for microbiology and other *in vitro* culture-based experiments.

Drawing connections and distinctions between the reported ptychographic sensor and other microscopy techniques helps to clarify and summarize the advantages of our device. Multi-height and multi-wavelength lensless imaging introduces multi-height and multi-wavelength diversity measurements for the phase retrieval process (Bao et al. 2008; Greenbaum and Ozcan 2012; Greenbaum et al. 2014; Latychevskaia 2019; Luo et al. 2016; Wu et al. 2021). The field of view is often limited to the area of the sensing surface, typically 20 – 40 mm$^2$. Without aperture synthesizing, the best-achieved resolution is similar to that of the ptychographic sensor. However, it is challenging for these techniques to restore the correct phase wraps of complex objects. Compared with these lensless techniques, ptychographic sensor employs transverse translation diversity measurements for the phase retrieval process. The lateral scanning of our device naturally expands the field of view beyond that is limited by the sensor size. As shown in Supplementary Fig. S6, we can image a 120-mm$^2$ field of view at a high spatiotemporal resolution. Another key advantage of the ptychographic sensor is the recovery of phase wraps of the bacterial colonies on uneven agar plates. Combined with the proposed temporal constraint, this unique capability allows us to quantitively track bacterial growth in a 15-second interval (Supplementary Video S1). For computational complexity, the ptychographic sensor shares a similar workload as other iterative phase retrieval approaches including multi-height and multi-wavelength lensless imaging. The most time-consuming operation in all these approaches is the fast Fourier transform (FFT) of the large-scale images. In our implementation, the processing time for 70 raw images with 2048 by 2048 raw pixels each is ~2 mins for 3 iterations using a Dell Alienware Aurora desktop. It is also possible to use parallel programming platforms such as Compute Unified Device Architecture (CUDA) to significantly shorten the processing time.

Lensless contact imaging and optofluidic microscopy, on the other hand, does not rely on phase retrieval for reconstruction (Cui et al. 2008; Jung and Lee 2016; Lee et al. 2014; Lee et al. 2012; Zheng et al. 2011; Zheng et al. 2010). The captured images are simply intensity projections of the cells. The imaging field of view is strictly limited by the size of the sensor and the achieved best resolution is at sub-micron level when applying the pixel super-resolution technique (Zheng et al. 2011; Zheng et al. 2010). The main drawback of these approaches is the close-contact requirement between the specimen and the sensing surface. They cannot image biospecimen cultured on regular Petri dishes. The ptychographic sensor, on the other hand, is fully compatible with existing culture-based experiments without the close-contact requirement. In the current implementation, we use the ptychographic sensor to directly acquire images through the agar plate. In other words, we have an agar layer, a substrate layer for agar plate, a coverglass layer of the sensor, and some additional free spaces between the specimen and the sensor surface.

Conventional ptychography uses a confined probe beam for object illumination. By translating the object (or the probe) to different lateral positions, it acquires the corresponding diffraction measurements for reconstruction (Faulkner and Rodenburg 2004). In the optical region, the technology has been commercialized using a lens-based setup (Godden et al. 2016). For a 3D thick sample such as a bacterial colony, however, the interaction between the confined beam and the sample cannot be modeled by a simple multiplication process (Zheng et al. 2021). The 3D nature of both the sample and the probe beam needs to be considered in the forward acquisition model. The reported ptychographic sensor, in contrast, performs modulation at the detection path. In this case, the recovered complex image represents the exit wavefront of the object. The thickness of the object is irrelevant in the modeling. The use of the unconfined scattering layer further improves the throughput.

Conventional lens-based microscope platforms can image bacterial cells in high-resolution, with a numerical aperture exceeding 1 in water. However, it is challenging to monitor bacterial growth during a culturing experiment using these platforms. For an up-right microscope, one need to take away the Petri dish cover to get close contact with the bacterial cells. For an inverted microscope, imaging through the agar layer is often impossible. Autofocusing is another major issue when imaging a large area of agar plate through scanning (Bian et al. 2020). For the ptychographic sensor, on the other hand, we can directly place the intact Petri dish on top of the sensor for image acquisition. Post-acquisition autofocusing can be performed by propagating back the recovered exit waves of the specimen. Compared with the regular microscope platform, a major disadvantage of the ptychographic sensor is, perhaps, the coherent nature of the imaging process. It cannot be used to image fluorescence signal in the current setting. We note that, however, it is possible to employ translated speckle illumination for modulating the fluorescence signal for lensless detection (Guo et al. 2018).

There are several future developments and applications of the ptychographic sensor. First, our device is compact and at a low cost. It can be fit within a small incubator and made broadly available for researchers. It provides a turnkey solution for performing longitudinal studies in bioscience research. In clinical applications, this can significantly cut down on the diagnostic time for medical procedures that requires culture growth-based assessment. For example, the ptychographic sensor can be used for tuberculosis, staph, and other bacterial infection diagnosis. The device can continuously and autonomously monitor for growth changes and notify the user to examine the sample when changes have been detected. Second, the illumination of the current device is based on one plane wave with a fixed incident angle. It is possible to perform synthetic aperture imaging using angle-varied illumination, similar to that demonstrated in Fourier ptychography (Zheng et al. 2013) and multi-wavelength phase retrieval (Luo et al. 2016). For imaging 3D bacterial colonies on solid agar plate, however, we need to consider the 3D nature of the sample. Each captured image can be used to update a spherical cap of the Ewald sphere in the 3D Fourier spectrum, as that demonstrated in Fourier ptychographic diffraction tomography (Horstmeyer et al. 2016; Zuo et al. 2020). Third, the design the of scattering layer on the sensor chip has not been optimized in the current device. It is possible to further improve the resolution with an optimal design of the layer, with some similarities to the design of metasurface. Fourth, the reported device is a platform technology. It is straightforward to integrate the device with a large number of sophisticated lab-on-a-chip designs, such as microorganism detection based on dielectrophoretic cages (Medoro et al. 2003), droplet-based platforms for cell encapsulation and screening (Clausell-Tormos et al. 2008), microfluidics-based phenotyping imaging and screening of multicellular organisms (Crane et al. 2010), high throughput malaria-infected erythrocyte separation and imaging (Hou et al. 2010), and screening of circulating tumor cells in high throughput (Williams et al. 2014; Zheng et al. 2019). Fifth, the ptychographic sensor can also be used with other microscopy techniques for correlative imaging. For example, it can be used with conventional bright-field or confocal microscopy for sample pre-screening and alignment. It can also locate individual filamentous cells over a large field of view for further analysis using other techniques. Lastly, the reported device provides quantitative phase imaging capability with a high throughput. A wide range of biomedicine-related applications (Park et al. 2018) can benefit from the device.

**CRediT authorship contribution statement**
**Shaowei Jiang**: Methodology, Validation, Investigation, Writing – original draft, Visualization. **Chengfei Guo**: Methodology, Validation, Investigation, Writing – original draft, Visualization. **Zichao Bian**: Methodology, Validation, Investigation. **Ruihai Wang**: Methodology, Validation, Investigation. **Jiakai Zhu**: Methodology, Validation, Investigation. **Pengming Song**: Methodology, Validation, Investigation. **Patrick Hu**: Validation, Investigation. **Derek Hu**: Validation, Investigation. **Zibang Zhang**: Investigation, Writing – review & editing. **Kazunori Hoshino**: Methodology, Writing – review & editing. **Bin Feng**: Methodology, Writing – review & editing. **Guoan Zheng**: Conceptualization, Methodology, Resources,

Supervision, Validation, Writing – original draft, Writing – review & editing, Funding acquisition, Project administration.

**Declaration of competing interest**

The authors declare that they have no known competing financial interests or personal relationships that could have appeared to influence the work reported in this paper.

**Acknowledgment**

This work was supported partially by National Science Foundation 2012140 and the UConn SPARK grant. P. S. acknowledges the support of the Thermo Fisher Scientific fellowship.